\documentclass[apjl,iop,pdftex]{emulateapj}

\usepackage{epstopdf}  
\usepackage{ifthen}
\usepackage{rotating}
\usepackage{subfigure} 
\usepackage{booktabs}
\usepackage{footnote}

\usepackage{appendix}
\usepackage{color}

\submitted{Accepted for publication in Philosophical Transactions A, as part of the special issue ``Cometary Science after Rosetta''}
\shorttitle{SOHO Comets: 20-Years and 3,000 Objects Later}
\shortauthors{Battams and Knight}

\begin{document}
\bibliographystyle{apj}

\title{SOHO Comets: 20-Years and 3,000 Objects Later}

\author{Karl Battams\altaffilmark{1,2}, Matthew M. Knight\altaffilmark{3}}

\altaffiltext{1}{Contacting author: karl.battams@nrl.navy.mil}
\altaffiltext{2}{US Naval Research Laboratory-Code 7685, 4555 Overlook Avenue, SW, Washington, DC 20375, USA}
\altaffiltext{3}{Department of Astronomy, University of Maryland, 1113 Physical Sciences Complex, Building 415, College Park, MD 20742, USA}

\begin{abstract}
We present a summary of the more than 3,000 sungrazing and near-Sun comets discovered in coronagraph images returned by the Solar and Heliospheric Observatory ({\it SOHO}), since its launch in December 1995. We address each of the four main populations of objects observed by {\it SOHO}: Kreutz (sungrazing) group, Meyer group, Marsden and Kracht (96P-Family) group, and non-group comets. Discussions for each group include basic properties, discovery statistics, and morphological appearance. In addition to updating the community on the status of the discoveries by {\it SOHO}, we also show that the rate of discovery of Kreutz sungrazers has likely remained static since approximately 2003, and report on the first likely fragmentation pair observed within the Meyer group.
\end{abstract}


\section{INTRODUCTION}
The joint ESA-NASA Solar and Heliospheric Observatory ({\it SOHO}) was launched in late 1995 and began routine operations in April of 1996. Since that time, and by virtue of the unique views offered by the on-board Large Angle Spectrometric Coronagraph (LASCO) instruments, more than 3,000 previously unknown sungrazing and near-Sun comets have been discovered in {\it SOHO} images. The overwhelming majority of {\it SOHO} discoveries have been made by amateur astronomers and enabled by the NASA-funded Sungrazing Comets Project, which was initiated in 2000 in response to increased public awareness of the {\it SOHO} data.

{\it SOHO}'s first comet discovery -- Kreutz comet C/1996~Q2 (SOHO) -- was made on August 22, 1998, by mission team member S.\ Stezelberger, though the publication of the orbit was not made until the following year \citep{stcyr97}. Over the next two years, several bright comets were detected in {\it SOHO}'s LASCO cameras by {\it SOHO} mission team scientists. 

With word slowly reaching the broader astronomical community, amateur astronomers not affiliated with the {\it SOHO} mission began to take advantage of {\it SOHO} data posted publicly to the Internet, and  began reporting comets in archive data previously overlooked by scientists. By 2000, approximately two hundred comets had been detected in {\it SOHO}'s LASCO imagery.

The increasing public awareness of this led directly to the creation of what is now known as the ``Sungrazer Project''\footnote{http://sungrazer.nrl.navy.mil/} -- the NASA-funded citizen science effort that enables amateur astronomers and enthusiasts worldwide to search for and report comets in {\it SOHO} data. This project was curated by D.\ Biesecker from its inception through 2003, D.\ Hammer for part of 2003, and then by K.\ Battams from October 2003 to the present time. During this period, over 3,100 new objects have been discovered via the Project.

In this paper we first outline the {\it SOHO}/LASCO observations and instrument specifications, and give an overview of spacecraft operational changes that have occurred during the mission. Following a brief definition of the terminolgy we use in the paper, we then discuss each of the four primary populations of objects observed and discovered by {\it SOHO}. These discussions provide a brief introduction to each population and cover the discovery statistics and typical morphologies of the objects. We show that the discovery rates of the most populous groups -- Kreutz and Meyer -- have remained largely static over at least the past decade or more.

\section{LASCO OBSERVATIONS}
{\it SOHO} carries 16 instruments designed to study the Sun from a position that orbits the Earth-Sun L1 Lagrange point \citep{domingo95}. Comets are discovered and observed primarily by LASCO, which itself is a suite of three coronagraph telescopes -- C1, C2, and C3 -- designed to observe the solar corona from a distance of 1.1--30 R$_\odot$ (1 R$_\odot$ = 1 solar radius = 0.00465 AU) in the plane of the sky \citep{brueckner95}. We note briefly that several comets have been observed by {\it SOHO}'s Ultraviolet Coronal Spectrograph (UVCS) (e.g., \citealt{raymond98}, \citealt{bemporad07}), and that comets are routinely observed and occasionally discovered in its all-sky Solar Wind Anisotropies (SWAN) Lyman-alpha imager (e.g., \citealt{bertaux97}, \citealt{combi14}). However, these observations and detections are beyond the scope of this paper.

LASCO's innermost field of view was covered by the internally-occulted Fabry-Perot interferometer C1 telescope that observed the corona from 1.1 to 3.0 R$_\odot$ at 5.6-arcsec/pixel resolution. No comets were detected by C1, which was only operational until June 24 1998, and thus no further mention of the instrument is made here. 

The LASCO C2 telescope is an externally-occulted coronagraph with a field of view extending from approximately 2.0 to 6.4 R$_\odot$. Images are recorded by a 1024$\times$1024 pixel CCD, with a resolution of 11.9 arcsec/pixel. Nominally, C2 images are recorded using a combination of Orange filter and Clear polarizer, the specifications of which are detailed in Table~\ref{t:filters}. Other filters available to C2 are Clear, Blue, Deep Red, H-alpha and Infrared, with polarizers at 0$^\circ$ and $\pm$60$^\circ$. Daily sequences of filtered and polarized images are recorded at half-resolution, or more frequently during dedicated observing campaigns, but the majority of C2 images are recorded with the Orange-Clear filter combination, with the H-alpha and Infrared filters having been used very rarely throughout the mission. Prior to 2010, C2 nominally returned three full-resolution images per hour, with that rate increasing to five per hour after 2010. The limiting magnitude for  C2 is approximately m9\footnote{{\it SOHO} magnitudes are calculated as roughly equivalent to Johnson V \citep{llebaria06, thernisien06}. Throughout this manuscript we refer to them with a leading ``m.''}, however the use of the orange filter means that this limit is influenced by the color of the object in question. The majority of {\it SOHO}'s comets appear brighter in C2 than C3, presumed due to the strong response of the LASCO orange filter to sodium emission.

\begin{table*}
\caption[LASCO filters and polarizers]{Typical filter and polarizer information for 2010--2016}
\vspace{-0.15in}
\begin{center}
\setlength{\tabcolsep}{0.05in}
\begin{tabular}{lccccc}
\hline
Filter&Telescope(s)&Bandpass&Image Size\footnote{Typical values; there has been some variation throughout the lifetime of {\it SOHO}}&Exposure Time$^\mathrm{a}$&Frequency$^\mathrm{a}$\\
&&(nm)&(pixels)&(sec)&(day$^{-1}$)\\
\hline
Clear&C3&400--850&1024$\times$1024&19 or 17&96--120\\
Blue&C2, C3& 420--520&512$\times$512&150 or 300&1\\
Orange&C2&540--640&1024$\times$1024&25&96--120\\
Orange&C3&540--640&512$\times$512&90 or 300&1\\
H$\alpha$&C2, C3&655--657&1024$\times$1024&300&rarely\\
Deep Red&C2, C3&730--835&512$\times$512&25 or 60&1\\
Infrared&C3&860--1050&512$\times$512&180&1\\
Polarizers (0$^\circ$, $\pm$60$^\circ$)&C2, C3&400--850&512$\times$512&100 or 300&1 sequence\\
\hline
\end{tabular}
\end{center}
\label{t:filters}
\end{table*}

The LASCO C3 telescope is an externally-occulted coronagraph with a field of view extending from approximately 3.7 to 30.0 R$_\odot$. Images are recorded by a 1024$\times$1024 pixel CCD, with a resolution of 56.0 arcsec/pixel. Nominally, C3 images are recorded using a combination of Clear filter and Clear polarizer, the specifications of which are detailed in Table~\ref{t:filters}, though the instrument has available the same filters as C2. Daily sequences of filtered and polarized images are recorded at half-resolution, or more frequently during dedicated observing campaigns, but the majority of C3 images are recorded with the Clear-Clear filter combination. Prior to 2010, C3 nominally returned three full-resolution images per hour, with that rate increasing to five per hour after 2010. The limiting magnitude for  C3 is approximately m8, however once again this limit is influenced by the color of object in question. 

In {\it SOHO}'s early mission, LASCO data coverage was significantly less than current rates. LASCO began recording images in January 1996, but the instrument was not commissioned until April of that year. Many images recorded during the first few months were not suitable for comet discovery, as the data rates were infrequent, data coverage at time sparse, and image sizes and exposure times variable. In 1998, the mission was interrupted from June until October when an erroneous command resulted in loss of communication with the satellite, and then again from December 1998 to February 1999. Smaller mission interruptions -- primarily planned, and typically one to three days in duration -- have occurred since then, but with lesser impact on comet detections.

In 2010, the {\it SOHO} mission underwent a significant programmatic shift, beginning to operate in an extended phase of the mission that took it outside of the scope of typical NASA heliophysics missions \citep{fletcher12}. This significantly de-scoped extension to the mission saw emphasis shift almost exclusively to LASCO coronagraph imagery. Accordingly, most instrumentation aboard {\it SOHO} was reduced to minimal telemetry or entirely ceased, with the additional bandwidth going to LASCO. As of August 3, 2010, both LASCO C3 and C2 cameras were increased to a nominal 12-minute cadence, with on-board recording of data through all major DSN gaps. LASCO is safed (i.e. no telemetry) for a few hours every three months when the {\it SOHO} spacecraft performs station keeping and momentum management maneuvers, but otherwise enjoys 24-hours data coverage at the highest cadence possible within the constraints of a twenty-plus year old satellite.

The increased data rate has seen a corresponding jump in the detection rates of {\it SOHO} comets, with approximately fifty more comets per year discovered during the extended phase. Primarily, the number of Kreutz-group comets detected in LASCO C2 comprise the bulk of these new discoveries. In each comet group section in this paper we discuss the impact of {\it SOHO} data rate upon discovery rates for those groups.

\section{Terminology}
\label{sec:terminology}
All objects discussed in this paper, whether members of known populations or sporadic single-apparition or short-period objects, will be referred to as comets. However, we apply this label despite minimal morphological support for categorizing them as such. The majority of {\it SOHO} ``comets'' have only been observed by the {\it SOHO} coronagraph instruments. It is only by the nature of the brightening observed around perihelion (or minimum elongation) that we infer the presence of a dust coma surrounding the object, and thus some form of physical activity at the object's nucleus. The brightness profiles observed by {\it SOHO} comets transiting the field of view would require bare nuclei of the order 10~km should they be inactive. While this can not be ruled out as a possibility, it seems unlikely that dozens or hundreds of kilometer-class objects would be repeatedly missed by modern sky surveys. However, we also note that the short perihelion distances of these objects expose them to solar radiation sufficient to sublimate many hard, non-volatiles (e.g., Table II of \citealt{mann04}), and thus even a small asteroidal object would likely present signs of physical activity near the Sun. It is likely that the {\it SOHO} non-Kreutz comet population comprises a mixture of asteroidal (``hot rocks'', cf. 3200 Phaethon; \citealt{jewitt10}, \citealt{li13b}) and exhausted cometary nuclei, in addition to ordinary cometary nuclei -- an issue explored by \citet{knight16} with regards to comet 322P/SOHO~1. 

We follow the convention of \citet{knight13b} that categorizes comets with perihelion distance inside the Sun's fluid Roche limit as ``sungrazing.'' For typical cometary densities of $\sim$500 kg m$^{-3}$ (e.g., \citealt{ahearn11}, \citealt{sierks15}) this is approximately 3~$R_\odot$ (0.014 AU). Thus we use the term ``sungrazing'' to encompass all known Kreutz objects and a small number of non-group {\it SOHO} comets. We use the term ``sunskirter'' to describe all other comets with perihelion distances within the fields of view of LASCO, e.g., the Meyer group, the 96P-family objects, and most non-group comets.

The terms ``C2'' and ``C3'' are used to refer respectively to the LASCO C2 and C3 coronagraphs aboard the {\it SOHO} satellite and  not to the common cometary emission bands C$_2$ and C$_3$.

\section{Summary of Discoveries}
\label{sec:disc}
As of May 31, 2016, 3,138 new comet discoveries are credited to {\it SOHO}. Of these, approximately 2,000 have official designations and orbits assigned by the Minor Planet Center (MPC), approximately 970 have astrometric reductions awaiting designation by the MPC, and the remainder are recent discoveries ``confirmed'' by the Sungrazer Project PI (Battams) and awaiting the final LASCO instrument data product before astrometric measurements are performed. These numbers include a few tens of periodic objects that will have been counted more than once on repeat apparitions, but due to the poor-quality of astrometric measurements obtained from {\it SOHO} data are unable to be definitely linked to one another. Once an object is identified definitively as periodic, future apparitions are not counted as additions to the total discovery count, but neither is the total count of discovered comets adjusted down. Twelve comets were discovered by the Solar Wind Anisotropies (SWAN) Lyman-alpha all-sky imager aboard {\it SOHO}, and carry the name ``SWAN.'' The remaining comets were all discovered by one of the two operational LASCO coronagraphs, C2 and C3 and carry the name ``SOHO.'' 

Table~\ref{table_overview} summarizes {\it SOHO}'s comet discoveries. Approximately 86\% are objects belonging to the extensive Kreutz group, the only family of near-Sun comets known to exist prior to {\it SOHO}. The Meyer group comprises the second most-populous group of objects, accounting for just over 6\% of {\it SOHO} discoveries. The Marsden and Kracht groups, with 54 and 42 members, respectively, together account for 3\% of {\it SOHO} objects. As will be discussed in Section~\ref{sec:96p}, these two families of short-period sunskirting comets are thought to be linked to the much larger comet 96P/Machholz~1, as part of the extended ``96P Family.'' Many of the Marsden and Kracht objects counted in column two of Table ~\ref{table_overview} will be repeat apparitions. The Kracht~II group contains just eight listed members, though likely comprises only two or three distinct objects, most notable comet 322P/SOHO~1.\footnote{While the MPC no longer assigns numbers for returning periodic comets of the same name, we use the numbers maintained by T.\ Farnham (http:// pdssbn.astro.umd.edu/data\_sb/resources/periodic\_comets.shtml) here to aid in distinguishing between objects.} Note that there is no dynamical linkage between the Kracht and Kracht~II groups; they are so named because both were first recognized by R. Kracht.  Finally, the ``non-group'' classification accounts for almost 5\% of {\it SOHO}'s discoveries, encompassing all objects that do not fall into one of the known populations. None of the non-Kreutz SOHO objects were known to exist prior to the {\it SOHO} mission, and only a few have been observed from the ground (e.g., C/1998~J1 SOHO, 322P/SOHO~1).
We discuss each of these groups in detail in the later sections of this paper, summarizing their discovery statistics and discussing their typical morphological characteristics. 
 
\begin{table*}
\begin{center}
\caption{Summary of {\it SOHO} Comet Groups, January 1996--May 2016}
\label{table_overview}
\begin{tabular}{lccccccc}
\hline
Group & Number & $q$ (AU)\footnote{The values given for orbital elements are averages of all values for those groups.} & $e$\footnote{Due to short orbital arcs $e$ is 1.0 for all {\it SOHO}-discovered comets unless it observed on a second apparition.} & $\omega$ ($^\circ$)& $\Omega$ ($^\circ$)& $i$ ($^\circ$)& Period (yr) \\
\hline
Kreutz 		&2692  &0.0056  & $>$0.9999 &80.0	& \phantom{0}0.4	& 143.2	& 500--1000\\
Meyer 		&\phantom{0}200  	&0.036\phantom{0}  & 1.0\phantom{00} &57.4	& 73.1	& \phantom{0}72.6	& Unknown \\ 
Non-Group 	&\phantom{0}142  	&Many  & Many &Many	& Many	& Many	& N/A \\ 
Marsden 	&\phantom{00}54  	&0.048\phantom{0}  & 0.984 &24.2	& 79.0	& \phantom{0}26.5	& 5.30--6.10 \\
Kracht 		&\phantom{00}42  	&0.045\phantom{0}  & 0.984 &58.8	& 43.8	& \phantom{0}13.4	& 4.81--5.81 \\
Kracht II 	&\phantom{000}8  	&0.054\phantom{0}  & 0.978 &48.6	& \phantom{0}0.0	& \phantom{0}12.6	& 3.99 \\\hline
TOTAL		& 3138 	&  	&   &  	&   &   	&   \\\hline 
\end{tabular}
\end{center}
\end{table*}

\begin{table}
\caption{Overview of images and discovery rates}
\begin{tabular}{lcccc}
\hline
Year & C2 Images\footnote{Number of full-resolution images returned by the LASCO instrument per year} & C3 Images$^\mathrm{a}$& Kreutz rate\footnote{Rate of discovery per 1000 combined C3 and C2 images} & Meyer rate\footnote{Rate of discovery per 1000 C2 images}\\
\hline
1996 &\phantom{1}6690  &\phantom{1}6987 & 1.97 & 0.15 \\
1997 &13257  &\phantom{1}8492 & 3.26 & 0.60   \\ 
1998 &\phantom{1}9072  &\phantom{1}5527 & 4.86 & 0.33   \\
1999 &16517  &11723 & 2.80 & 0.24   \\
2000 &22049  &14297 & 2.34 & 0.27   \\
2001 &23157  &14670 & 2.46 & 0.39   \\
2002 &22695 &91428 & 2.95 & 0.40   \\
2003 &21194 &14634 & 3.66 & 0.33   \\
2004 &21224 &13156 & 4.22 & 0.33   \\
2005 &24535 &14519 & 3.64 & 0.53  \\
2006 &23811 &13357 & 3.79 & 0.25   \\
2007 &22744 &14276 & 4.02 & 0.48  \\
2008 &22860 &14332 & 3.66 & 0.52  \\
2009 &21281 &14298 & 4.19 & 0.33   \\
2010 &28960 &24083 & 3.58 & 0.45  \\
2011 &37986 &37089 & 2.61 & 0.32  \\
2012 &37766 &37407 & 2.54 & 0.48   \\
2013 &38839 &38488 & 2.55 & 0.33  \\
2014 &39108 &38724 & 2.08 & 0.43  \\
2015 &34502 &34182 & 2.64 & 0.64  \\\hline 
\end{tabular}
\label{table_lasco_images}
\end{table}

\begin{figure}
  \centering
  \includegraphics[width=88mm]{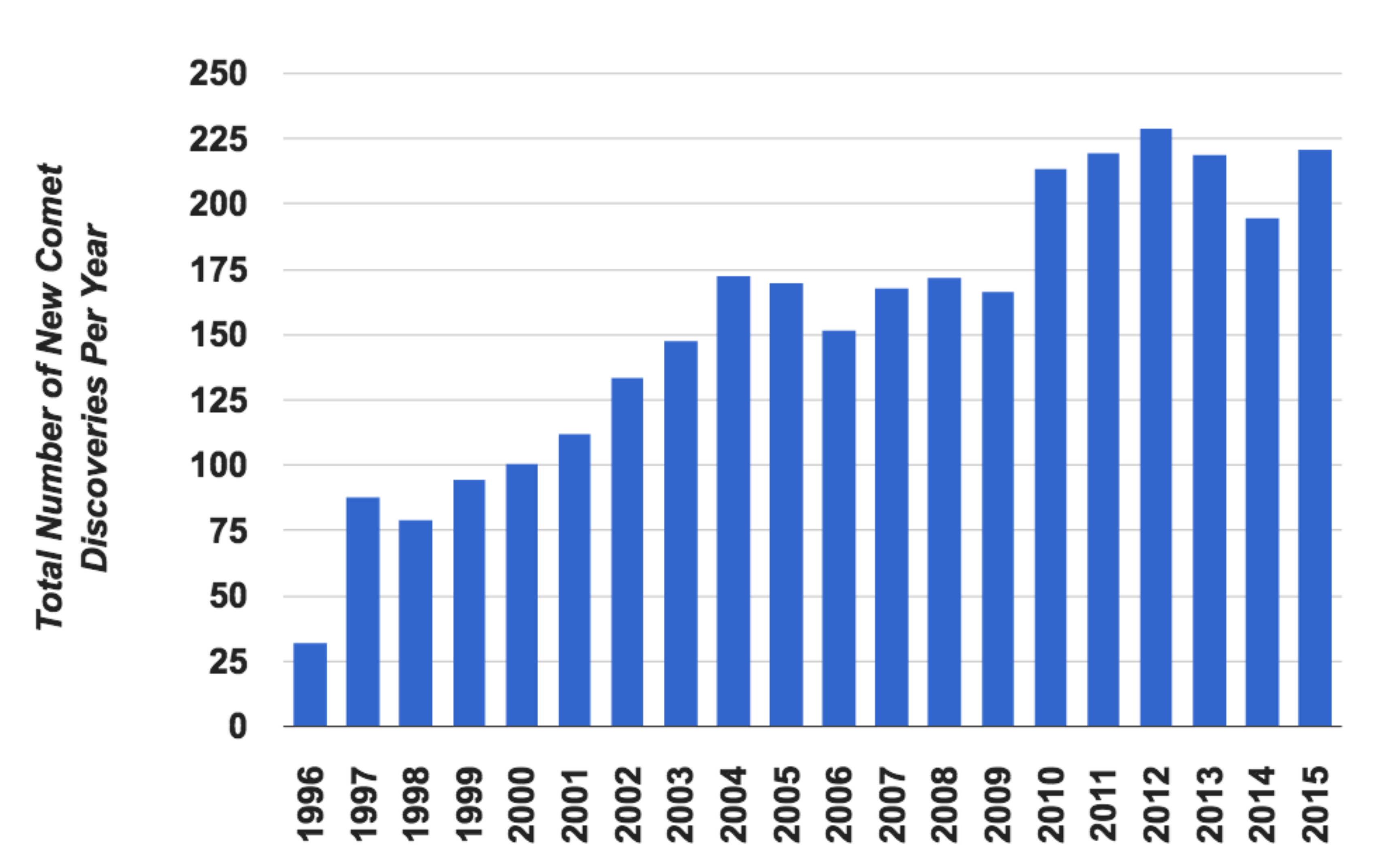}
  \caption[Images]{Total number of new {\it SOHO} comet discoveries per year from 1996 through 2015. }
\label{fig:all_discoveries}
\end{figure}

Table~\ref{table_lasco_images} shows the number of ``useful'' images returned by the LASCO C2 and C3 per year for 1996 through 2015, along with discovery rates for Kreutz and Meyer group comets for every 1,000 such images recorded. We define a ``useful'' image as a full-resolution 1024$\times$1024 pixel image using the Clear-Clear filter/polarizer in C3 and the Orange-Clear filter/polarizer in C2. A very small number of early-mission Kreutz were detected in 1024$\times$768 and 512$\times$512 images but such instances were rare, and only applicable to data recorded before 1998 when routine collection of images at those resolutions largely ceased. The Kreutz rates are based on the combined total of C2 and C3 images, whereas the Meyer rates are based solely on the number of C2 images as Meyer comets are rarely observed in C3. We analyze these numbers in more detail in the corresponding group sections later in this paper.

Figure~\ref{fig:all_discoveries} shows the total number of {\it SOHO} comet discoveries per complete calendar year since the mission began routine operations. By the end of December 2015, there were a total of 3,088 new comet discoveries made by the {\it SOHO} satellite. Notable in Figure~\ref{fig:all_discoveries} is an apparent steady annual increase in comet discovery rates until 2004, following which the rates remained largely constant until a large jump during 2010, to the present rate of approximately 220 new comet discoveries per year. 

The low discovery rate in 1996 is a consequence of the changeable and intermittent operations of the LASCO instrument during its first year of operation. The gradual annual increase in discoveries can be attributed in part to a stabilizing of the instrument telemetry rates and observing schedules as the mission progressed, although it has been suggested \citep{sekanina07,knight10} that there was an increase in the rate of Kreutz (which dominate detections) around 2003. The noticeable jump beginning in 2010 is attributed to the increase in telemetry for the LASCO instrument under the new mission operations scheme and will be discussed in more detail later.

Table~\ref{table_comets} breaks down the comet discoveries by year and by group. As noted earlier, the Kracht, Kracht~II, and Marsden group objects are known to be of short period and thus many objects listed are re-discoveries of prior objects that, either now or at the time, could not be linked to a previous apparition. We discuss this further in Section 5. There likely exist a small number of short-period comets within the Non-group categories, but again these are objects that can not be definitively linked to a prior apparition. New detections of ``known'' periodic comets are not counted in the statistics. All Kreutz and Meyer objects are assumed to be unique apparitions. 
\vfill 

\begin{table*}
\caption{Yearly {\it SOHO} Comet Discoveries}
\vspace{-0.15in}
\begin{center}
\label{table_comets}
\begin{tabular}{cccccccc}
\hline
Year\phantom{$^q$} & Kreutz & Meyer & Kracht & Kracht II & Marsden& Non-Group & Total \\
\hline
1996\phantom{$^q$} &\phantom{1}27  &\phantom{1}1  &\phantom{1}3  & 0 &1& \phantom{1}0& \phantom{1}32 \\
1997\phantom{$^q$} &\phantom{1}71  &\phantom{1}8  &\phantom{1}0  & 0 &3& \phantom{1}6& \phantom{1}88 \\ 
1998\phantom{$^q$} &\phantom{1}71  &\phantom{1}3  &\phantom{1}0  & 0 &3& \phantom{1}2& \phantom{1}79 \\
1999\phantom{$^q$} &\phantom{1}79  &\phantom{1}4  &\phantom{1}2  & 1 &6& \phantom{1}3& \phantom{1}95 \\
2000\phantom{$^q$} &\phantom{1}85  &\phantom{1}6  &\phantom{1}1  & 0 &3& \phantom{1}6& 101 \\
2001\phantom{$^q$} &\phantom{1}93  &\phantom{1}9  &\phantom{1}4  & 0 &0& \phantom{1}6& 112 \\
2002\phantom{$^q$} &109 &\phantom{1}9  &\phantom{1}7  & 1 &3& \phantom{1}5& 134 \\
2003\phantom{$^q$} &131 &\phantom{1}7  &\phantom{1}0  & 1 &2& \phantom{1}7& 148 \\
2004\phantom{$^q$} &145 &\phantom{1}7  &11 & 0 &3& \phantom{1}7& 173 \\
2005\phantom{$^q$} &142 &13 &\phantom{1}1  & 0 &4& 10& 170 \\
2006\phantom{$^q$} &141 &\phantom{1}6  &\phantom{1}0  & 0 &2& \phantom{1}3& 152 \\
2007\phantom{$^q$} &149 &11 &\phantom{1}0  & 0 &1& \phantom{1}7& 168 \\
2008\phantom{$^q$} &136 &12 &\phantom{1}4  & 2 &4& 14& 172 \\
2009\phantom{$^q$} &149 &\phantom{1}7  &\phantom{1}3  & 0 &3& \phantom{1}5& 167 \\
2010\phantom{$^q$} &190 &13 &\phantom{1}0  & 1 &3& \phantom{1}7& 214 \\
2011\phantom{$^q$} &196 &12 &\phantom{1}1  & 0 &4& \phantom{1}7& 220 \\
2012\phantom{$^q$} &191 &18 &\phantom{1}1  & 0 &6& 13& 229 \\
2013\phantom{$^q$} &197 &13 &\phantom{1}1  & 0 &0& \phantom{1}7& 218 \\
2014\phantom{$^q$} &162 &17 &\phantom{1}2  & 2 &1& 11& 195 \\
2015\phantom{$^q$}&181 &22 &\phantom{1}1  & 0 &2& 15& 218 \\
2016\footnote{Discoveries through May 31, 2016} &\phantom{1}47 &\phantom{1}2 &\phantom{1}0 & 0 & 0& \phantom{1}1 &\phantom{1}50 \\\hline
TOTAL& 2692& 200& 42 & 8 & 54& 142 & 3138\\\hline 
\end{tabular}
\end{center}
\end{table*}

\section{Comet Groups}
\label{sec:groups}
\subsection{Kreutz}
\label{sec:kreutz}
Prior to {\it SOHO}'s launch, approximately thirty Kreutz sungrazers were known. The group is named for Heinrich Kreutz who derived the first definitive linkages of very bright comets seen near the Sun in the 1800s \citep{kreutz88, kreutz91, kreutz01}.  With further Kreutz discoveries made in the mid-twentieth century (C/1945~X1 Du Toit, C/1963~R1 Pereyra, C/1965~S1 Ikeya-Seki and C/1970~K1 White-Ortiz-Bolelli), the group stood at nine confirmed members until 1979, when the SOLWIND coronagraph about the USAF P78-1/SOLWIND satellite observed the first space-detected comet -- the Kreutz group object C/1979~Q1 \citep{michels82a, sheeley82}. This discovery was followed by an additional eight likely Kreutz sungrazing comet detections by the SOLWIND instrument (including four archival discoveries by R. Kracht in 2005) before the satellite was destroyed by a planned U.S. Air Force exercise in 1985. The Solar Maximum Mission (SMM) coronagraph/polarimeter discovered an additional ten Kreutz comets (e.g., \citealt{macqueen91}) between 1984 and 1989. None of the comets discovered by SOLWIND and SMM were ground-observed and were thus likely similar to the brighter Kreutz comets observed by {\it SOHO}. These objects differ significantly in brightness and, presumably, size from the large ground-observed Kreutz of the nineteenth and twentieth centuries. For more detailed discussion of the history of the Kreutz group, we refer the reader to works such as \citet{kresak66}, \citet{marsden67,marsden89,marsden05}, and \citet{sekanina67b,sekanina67a,sekanina03}. 

Kreutz comets are characterized by their extremely small perihelion distances of around 1--2 $R_{\odot}$, high inclinations ($i{\sim}140^{\circ}$), and orbital periods of centuries. These orbital elements are primarily constrained from historic Kreutz observations; {\it SOHO}'s large pixel sizes and short observing arcs yield highly ambiguous orbits. Orbital elements stated for {\it SOHO}-observed Kreutz should be treated with the utmost of caution and skepticism and, in many cases, ``should not be taken at all seriously'' (Marsden, private comm.). This statement applies equally to all other {\it SOHO} discoveries other than those designated periodic.

The Kreutz lightcurves are unusual in that they generally peak in brightness at 10--15 $R_\odot$ and fade interior to this, with an occasional second peak inside of ${\sim}8~R_\odot$ \citep{biesecker02,knight10}. The comets eventually disappear, often significantly before perihelion, and none of the {\it SOHO}-observed Kreutz have been observed to survive perihelion except C/2011~W3 Lovejoy. This behavior is interpreted as the total destruction and vaporization of the comet and has been used to estimate initial sizes of less than $\sim$100 m in diameter \citep{iseli02,sekanina03,knight10}. The smallest Kreutz comets detected by {\it SOHO} are likely 5--10 m in diameter prior to the onset of activity; it is assumed that smaller objects exist but are below the detection threshold.

\subsubsection{Kreutz Morphology}
\begin{figure}
  \centering
  \includegraphics[width=88mm]{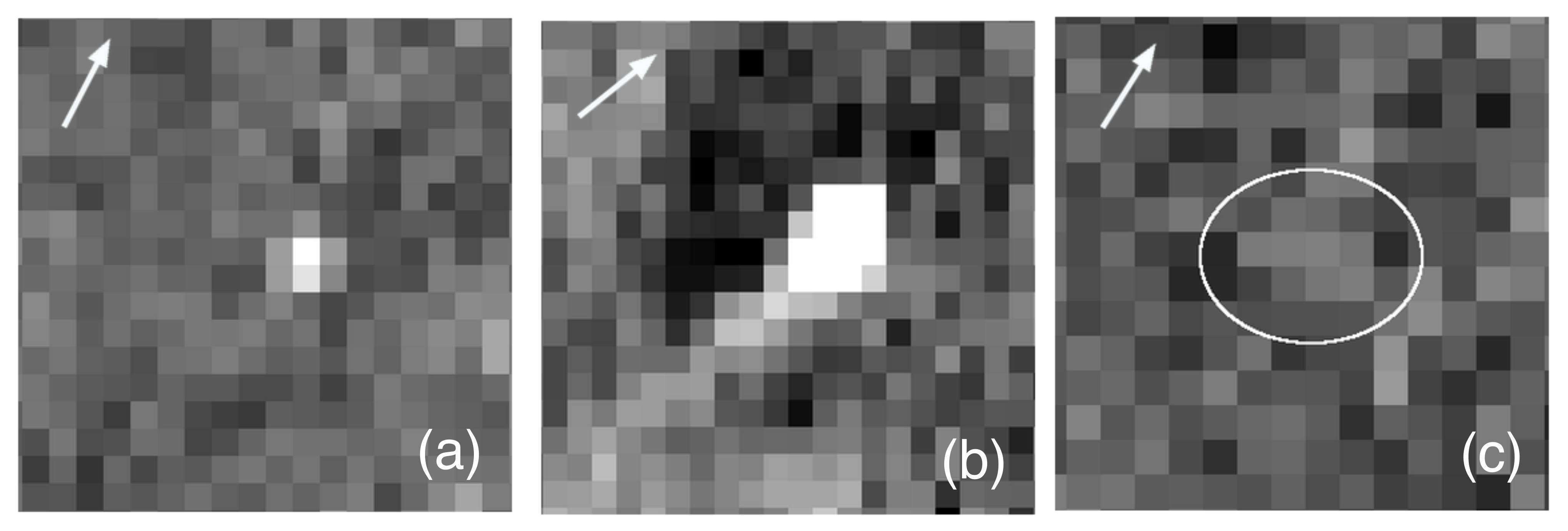}
  \caption[Images]{Typical morphologies of {\it SOHO}-observed Kreutz comets in LASCO C3 described as: (a) stellar; (b) stellar with tail; (c) diffuse, with object circled for easier identification. The white arrows point towards the Sun. In all cases, each pixel spans 56 arcsec. Object identifications are as follows: (a) SOHO-2871, Feb 10, 2015 15:06 UT; (b) SOHO-2865, Jan 26, 2015 14:18 UT; (c) SOHO-2869, Feb 09, 2015 04:30 UT. }
\label{fig:c3_krz}
\end{figure}

Unlike all other {\it SOHO}-observed populations, {\it SOHO}'s Kreutz comets exhibit diverse morphologies. We broadly categorize {\it SOHO}-observed Kreutz morphologies as follows: stellar (Figure~\ref{fig:c3_krz}a); tailed  (Figure~\ref{fig:c3_krz}b); and diffuse  (Figure~\ref{fig:c3_krz}c). Due to the large pixel scale (56 arcsec/pixel), C3 Kreutz comets rarely cover more than one to four pixels in any given image. Consequently, the majority of Kreutz observed in LASCO C3 are what we would describe as stellar; that is, very condensed with no apparent tail or diffuse extended coma. Kreutz brighter than approximately m5 tend to exhibit a tail feature in C3. 

\begin{figure}
  \centering
  \includegraphics[width=88mm]{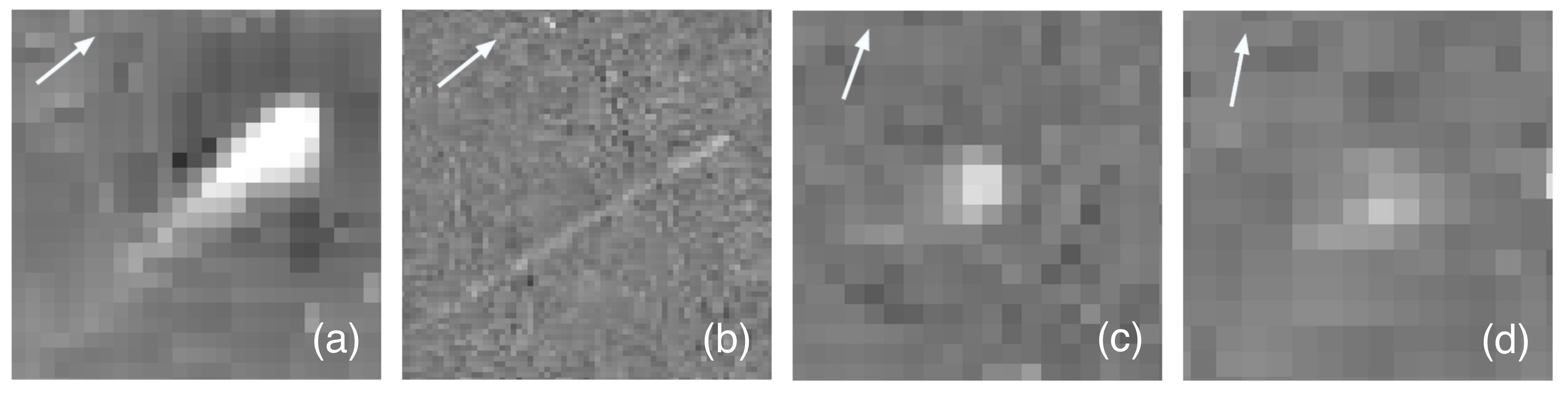}
  \caption[Images]{Typical morphologies of {\it SOHO}-observed Kreutz comets in LASCO C2 described as: (a) a tail and condensation; (b) only a tail (i.e. ``streak''); (c) condensed with little or no elongation; (d) diffuse. The white arrows point towards the Sun. In all cases, each pixel spans 11.9 arcsec. Object identifications are as follows: (a) SOHO-2878, Feb 23, 2015 20:24 UT; (b) SOHO-2880, Feb 24, 2015 20:00 UT; (c) SOHO-2897, Mar 29, 2015 04:00 UT; (d) SOHO-2915, Apr 12, 2015 11:00 UT. }
\label{fig:c2_krz}
\end{figure}

The improved 11.9 arcsec/pixel resolution offered by C2 enables a somewhat better qualitative analysis of the morphology. C2 Kreutz will typically span at least two pixels, and frequently as many as five or six. At this resolution, we routinely observe the objects' diffuse coma, and can also better detect tail-like features, even in faint comets. We also see a broader diversity of morphologies, including: small (one or two-pixel) stellar objects with no visible diffusivity or tail-like extension  (Figure~\ref{fig:c2_krz}c); small to large ($>$5 pixel) diffuse objects with no apparent central condensation  (Figure~\ref{fig:c2_krz}d); narrow, needle-like objects with no obvious head or nucleus  (Figure~\ref{fig:c2_krz}b); and classic cometary objects with coma, condensed central nucleus, and short tail  (Figure~\ref{fig:c2_krz}a). While these labels apply generally to all C2-observed Kreutz, the morphology often can evolve over time, and frequently overlaps between the categories we have defined.

\begin{figure}
  \centering
  \includegraphics[width=88mm]{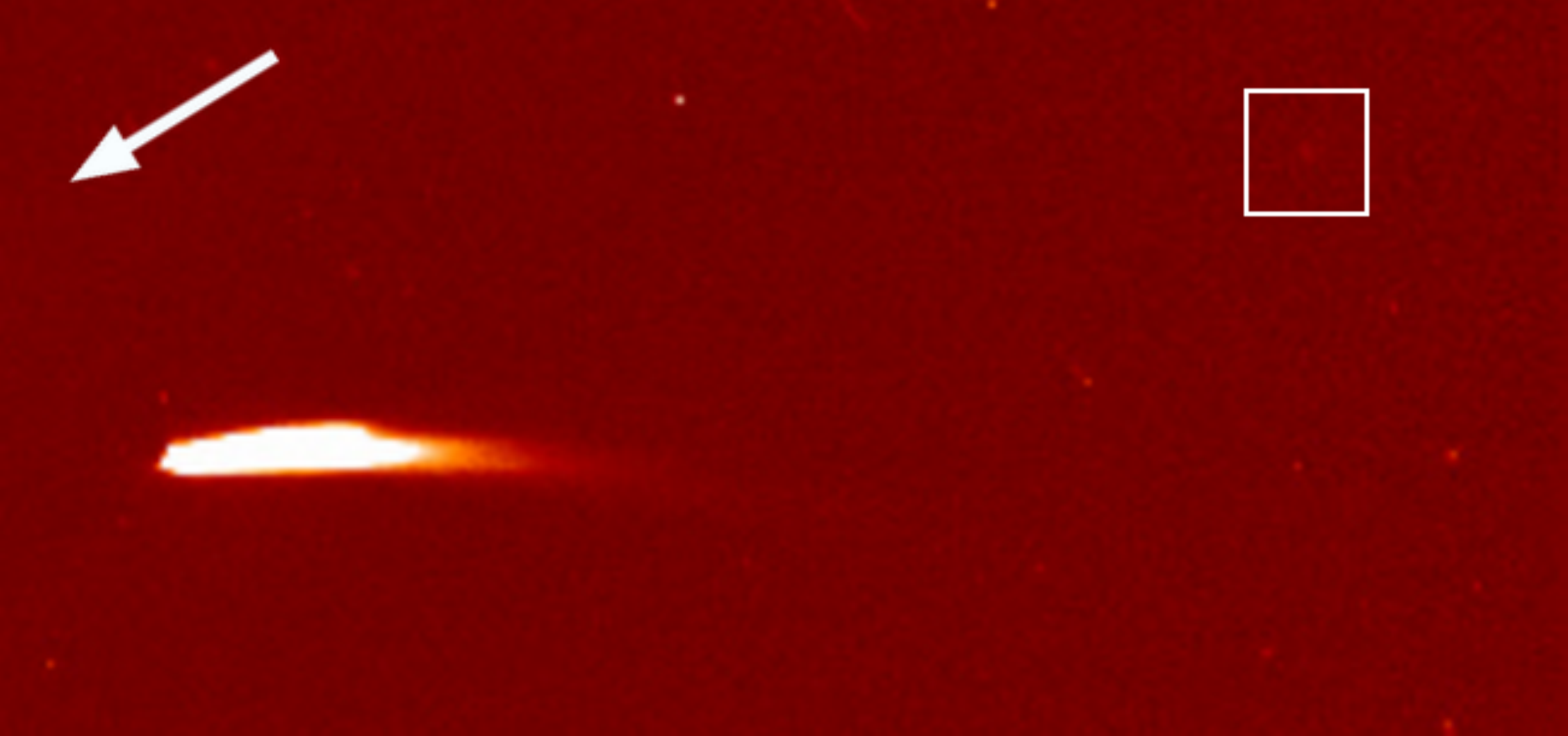}
  \caption[Images]{Comet C/2008 K4 (SOHO) trailed by comet C/2008 K5 (SOHO), indicated by the white box, at an apparent distance of approximately half a degree in the LASCO C2 field of view. The white arrow points towards solar north. The figure is taken from an image recorded on May 23, 2008 at 14:54 UT.}
\label{fig:fragment}
\end{figure}

To date, there has been no published study dedicated to categorizing {\it SOHO} Kreutz comet morphologies. While a time-consuming task, we feel it would be a potentially informative study. Such work could seek to filter out effects from seasonality (viewing geometry) and phase angle to produce a better understanding of the diverse apparent morphology of Kreutz sungrazers.

As discussed by \citet{sekanina00a,sekanina02a} {\it SOHO} Kreutz objects are prone to clustering, with one theory being that these are objects recently fragmented from one-another. However, the density of the Kreutz stream and ambiguities in orbit determinations make this a challenging case to definitively argue as there may be many serendipitously timed arrivals. On many occasions of two or more Kreutz passages within hours of each other, we have noted the morphologies of the two or more presumed siblings to be remarkably different, with a ``small diffuse'' Kreutz perhaps accompanying a much larger, condensed object, for example. Figure~\ref{fig:fragment} shows the very bright Kreutz comet C/2008 K4 (SOHO) trailed by approximately half a degree by the far smaller and fainter C/2008 K5 (SOHO), in an image recorded on May 23, 2008 at 14:54 UT. Numerous similar examples exist in the archives in which a bright object is followed or preceded by a very faint object. However, there also exist a substantial number of equivalently bright pairs of objects reaching perihelion within hours of each other. These pairs of comets have been studied by \citet{sekanina00a}; this is a subject that would benefit from revisiting with the additional sixteen years of available data.

\subsubsection{Kreutz Detection Rate}

\begin{figure}
  \centering
  \includegraphics[width=88mm]{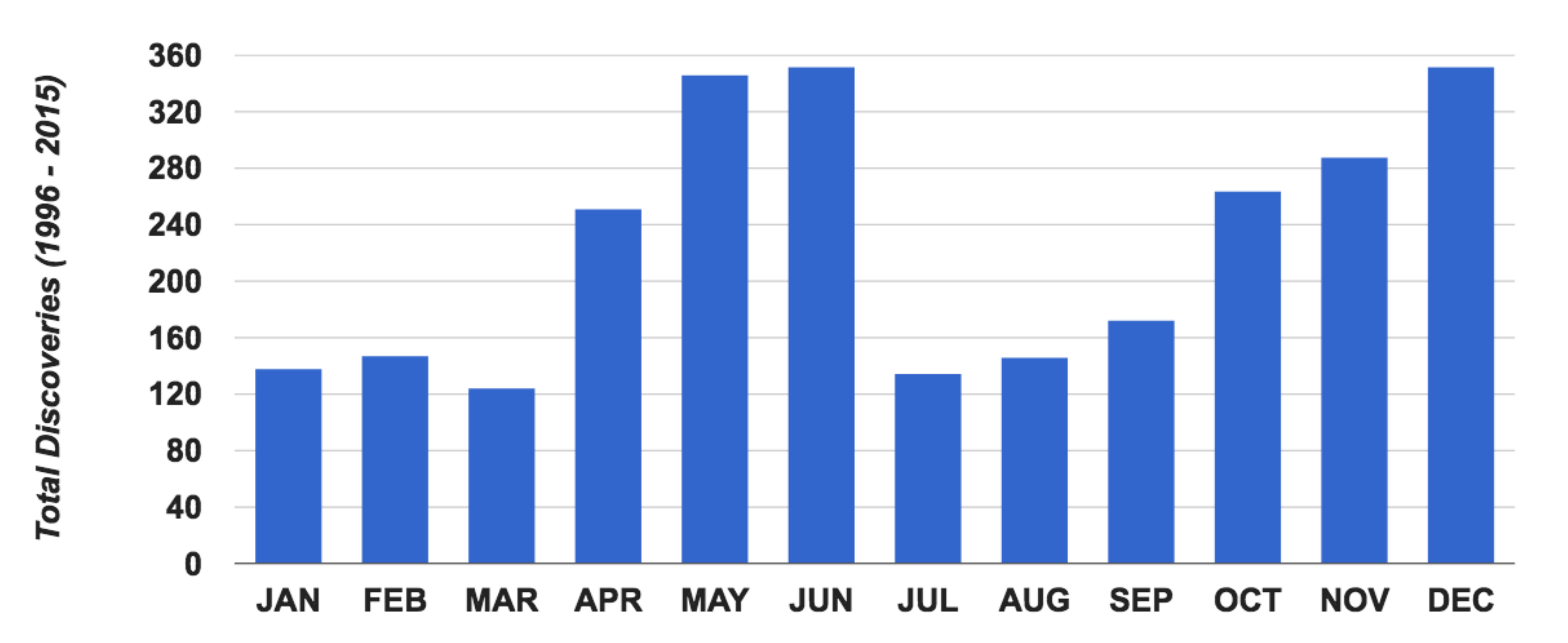}
  \caption[Images]{Total number of Kreutz discoveries in both C2 and C3 for each month, in the period 1996--2015. The April/May/June and October/November/December seasonal bias is due to more favorable viewing geometries of the Kreutz orbit resulting in many more objects detected in the C2 instrument.}
\label{fig:krz_seasons}
\end{figure}

Kreutz-comet discovery rates see significant impacts from the rate of data returned by LASCO, and seasonal effects resulting from the viewing geometry of the spacecraft. Figure~\ref{fig:krz_seasons} shows the total number of Kreutz discoveries in each month of the year for the entire {\it SOHO} mission, 1996--2015. The peaks in April/May/June and October/November/December correspond to the periods in which the geometry is such that the interval of peak brightness occurs in the more sensitive C2 camera. As discussed by \citet{knight10}, this seasonal variation is indicative that the actual rate of Kreutz comets is substantially higher than recorded. 

\begin{figure}
  \centering
  \includegraphics[width=88mm]{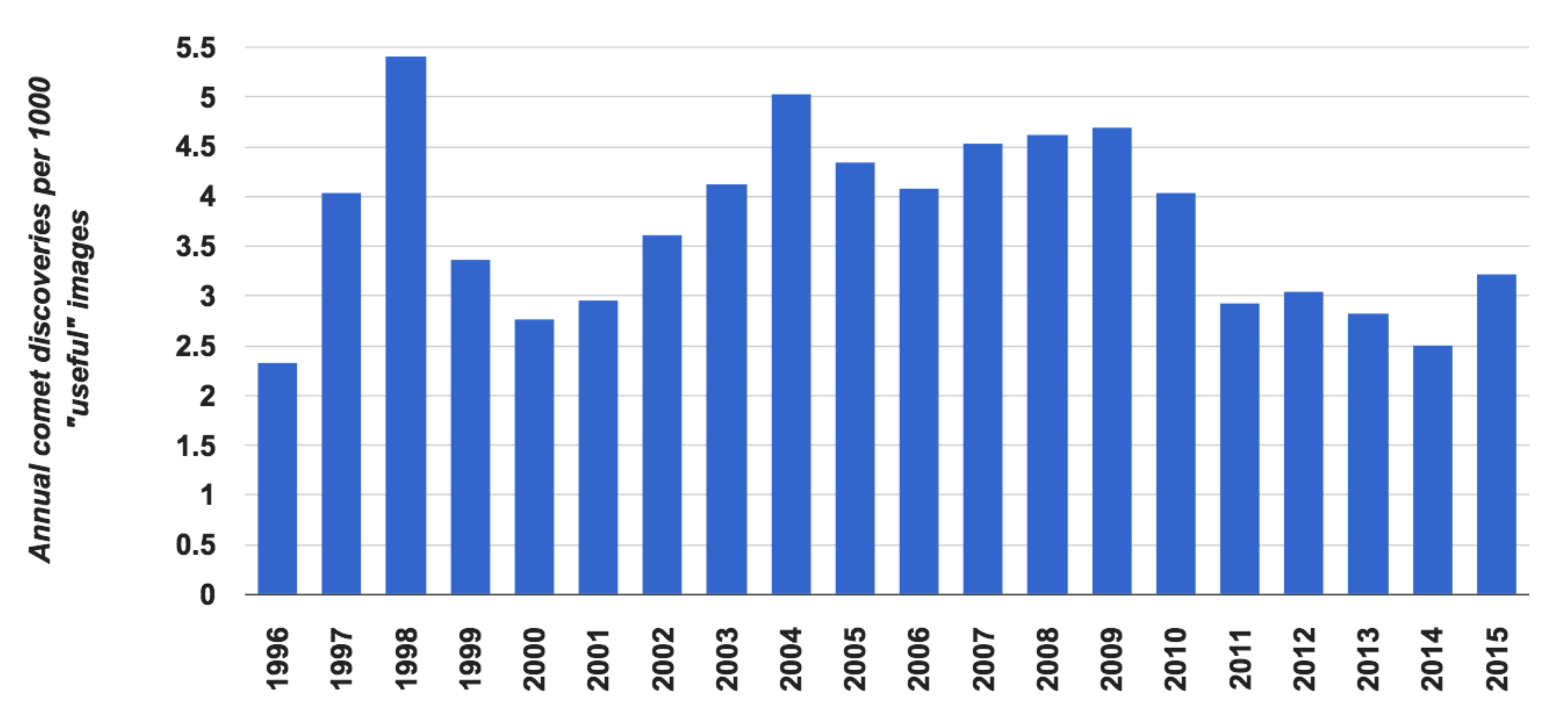}
  \caption[Images]{Yearly rate of Kreutz comet discovery per 1000 ``useful'' images recorded by the LASCO C2 and C3 cameras.}
\label{fig:krz_rates}
\end{figure}

The impact of the rate of data returned on {\it SOHO} Kreutz discoveries was considered by \citet{knight10}, who made estimates of the approximate number of comets likely ``missed'' due to observing interruptions. The study found that the rate of Kreutz comets reaching perihelion increased from 1996 to 2008, with a large jump occurring in 2002--2003, and that the increase was seen in comets at all sizes. Here we take an alternative approach and look at the rate of comet discovery normalized by the rate of ``useful'' images recorded annually by the spacecraft. The results are shown in Table~\ref{table_lasco_images}. We use the same definition of a useful image as outlined in Section~\ref{sec:disc}.

Figure~\ref{fig:krz_rates} plots the annual discovery rate of Kreutz comets per 1000 useful images recorded by both LASCO C2 and C3 cameras, as listed in column four of Table~\ref{table_lasco_images}. Over the past 19 years, the LASCO instrument has observed approximately three new Kreutz comets for every one thousand images returned by the instrument. 1998 shows the highest annual rate, with almost five per one thousand images. However this number must be treated with caution as the LASCO instrument was not operational through late June until late September of that year, which corresponds to the period in which fewer Kreutz comets tend to be observed, per Figure~\ref{fig:krz_seasons}. Thus this number is unfairly biased towards higher discovery rates.

Of particular note in Figure~\ref{fig:krz_rates} is the apparent fall in normalized discovery rates since 2010, despite a large increase in the total number of comets discovered during this period (Figure~\ref{fig:all_discoveries}). The years 2003 through 2010 show a very consistent normalized discovery rate of approximately 3.8 Kreutz comets per 1,000 useful images, whereas 2010 to present shows a similarly consistent rate of approximately 2.7 Kreutz comets per 1,000 useful images. In most cases, the additional comet discoveries since 2011 are extremely faint objects that may only be visible in five to eight consecutive images at the new higher data rate, and thus would have perhaps only been seen in two to four images during the mission's early phase. It has long been the policy of the Sungrazer Project that Kreutz comets could only be ``confirmed'' (and thus recorded) if at least five images of the comet are observed. This is to ensure that noise and cosmic rays are not falsely identified as comets, and that derived orbits of any objects have at least one hour of observing arc. LASCO C3 is less impacted by the increased data rate as objects discovered and observed in that camera tend to be those that are yet to peak, or have only just peaked in brightness, and thus generally persist for at least a few hours. Kreutz in C2, however, are often at the end of their life, frequently extremely small and likely vaporizing rapidly, thus five images per hour versus three per hour can have a major impact on their detectability. 

Two possibilities exist to explain the changes in discovery rates since 2010: (1) either the rate of Kreutz comets did indeed fall from constant levels in 2003--2010 to a new constant since 2011; or (2) we have reached a detection threshold based now on the sensitivity of the LASCO cameras and not on the rate of incoming comets. Since the number of comets per year has actually increased since 2010, the only explanation is (2). 

This implies that as the LASCO C2 data rate tended towards infinity, we would not detect appreciably more comets that are physically visible to the cameras at or above the instrument detection threshold. Thus, essentially all Kreutz objects that pass through the LASCO field of view that are at or above the detection threshold  are being discovered, i.e., the citizen scientists are not overlooking many objects. 

\subsection{Meyer}

As of May 31, 2016, 200 members of the Meyer-group had been identified, making this the second most populous known comet population. Yet despite their relatively high number, Meyer-group objects are the least well understood of all {\it SOHO} groups. Nothing is known of the origins or progenitor(s) of the group, and due to the low quality astrometry available from {\it SOHO}, their high-inclination orbit is only approximated. Thus, no reliable estimate of the orbital period is possible. Marsden (private comm.) has suggested the latter to be at least decades to centuries, while \citet{sekanina05} equivalently propose a likely large aphelion distance for this group. 

All observed members of the Meyer group transit LASCO C2, and with limited exceptions are exclusive to that instrument. Thus all discussion of Meyer comets in this paper should be assumed to refer to only C2 observations unless stated otherwise. Size estimates for Meyer comet are not possible, as all objects appear to survive perihelion so the assumptions made to estimate Kreutz sizes (e.g., \citealt{iseli02}, \citealt{sekanina03}, \citealt{knight10}) are not valid. Furthermore, none have been observed beyond the fields of view of {\it SOHO} (few are even seen in {\it SOHO}'s C3 field of view), so no useful constraints on size or activity level can be set at larger heliocentric distances. 

\begin{figure}
  \centering
  \includegraphics[width=88mm]{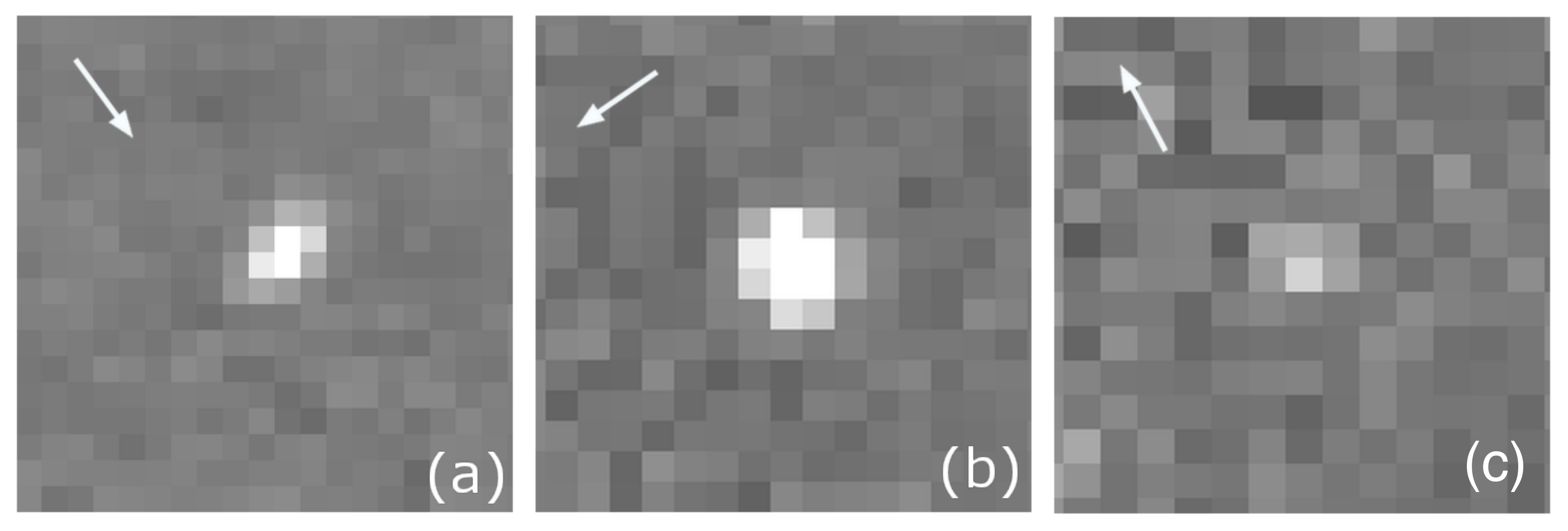}
  \caption[Images]{Typical morphologies of {\it SOHO}-observed comets in the LASCO C2 field of view belonging to the: (a) Meyer; (b) Kracht; and (c) Marsden groups. The white arrows point towards the Sun, and each pixel measures 11.9 arcsec in width. In all cases, and in all other observed instances of members of these groups, the comets are entirely stellar in appearance with no signs of a diffuse or extended appearance. Object identifications are as follows: (a) SOHO-2868, Jan 31, 2015 15:24 UT; (b) SOHO-2802, Sep 13, 2014 08:24 UT; (c) SOHO-2825, Nov 6, 2014 09:24 UT.}
\label{fig:non_krz_imgs}
\end{figure}

Morphologically, the Meyer comets are almost entirely uniform, exhibiting an identical stellar, condensed appearance. Meyer comets sometimes give the appearance of very slight elongation, but it remains unclear whether this is true elongation or simply an instrumental effect, as stars near the edge of the C2 field of view sometimes exhibit a similar apparent elongation. An example of a typical Meyer-group comet is shown in panel a of Figure~\ref{fig:non_krz_imgs}. The brighter members of the group reach approximately m6.5 \citep{lamy13}, but these are rare, with most objects around m7.5--8.5. Despite some inherently bright members of the population, none have been observed with any indication of a diffuse coma or obvious tail. 

Unlike Kreutz, Marsden and Kracht objects, the Meyer group shows little temporal clustering \citep{sekanina05}, perhaps implying they are structurally strong and/or have very low activity. The only notable ``cluster'' of Meyer-group comets on record are those of the recent trio, SOHO-2884, SOHO-2886 and SOHO-2887. The former of these was an unusually bright Meyer comet ($\sim$m6.5), observed in both LASCO C3 pre-and post-perihelion, as well as in LASCO C2, on March 1, 2015. The other two objects were observed in C2 only on March 5, 2015, following almost identical paths as each other and only two hours apart. This latter pairing constitutes the closest clustering of any Meyer group comets on record, and the distance between SOHO-2884 and 2886/2887 was only exceeded by the pair C/1997 U8 and U9 (SOHO), which were recorded approximately three days apart. Given the low rate of occurrence of Meyer comets (10 per year on average), the close clustering of SOHO-2886 and 2887 would imply a recent fragmentation event as opposed to an alternate hypothesis of simple coincidence. 

\begin{figure}
  \centering
  \includegraphics[width=88mm]{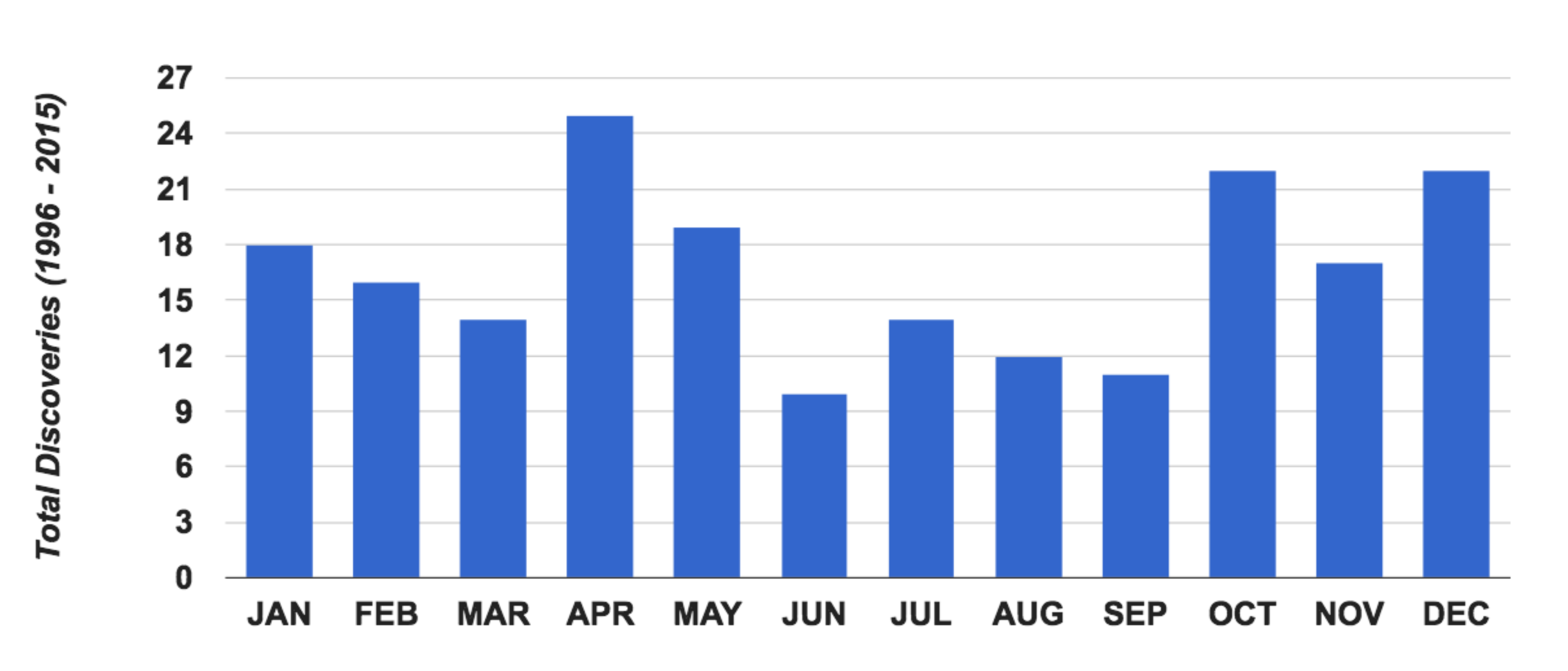}
  \caption[Images]{Total number of Meyer-group discoveries in each month during the period 1996--2015.}
\label{fig:meyer_seasons}
\end{figure}

The Meyer-group discoveries exhibit a small seasonal trend, as illustrated in Figure~\ref{fig:meyer_seasons}, but the effect is not as pronounced as that of the Kreutz group. Again, this seasonal effect is due to the viewing geometry of the Meyer-group orbits throughout the year, with April/May and October/November/December offering better geometries. 

\begin{figure}
  \centering
  \includegraphics[width=88mm]{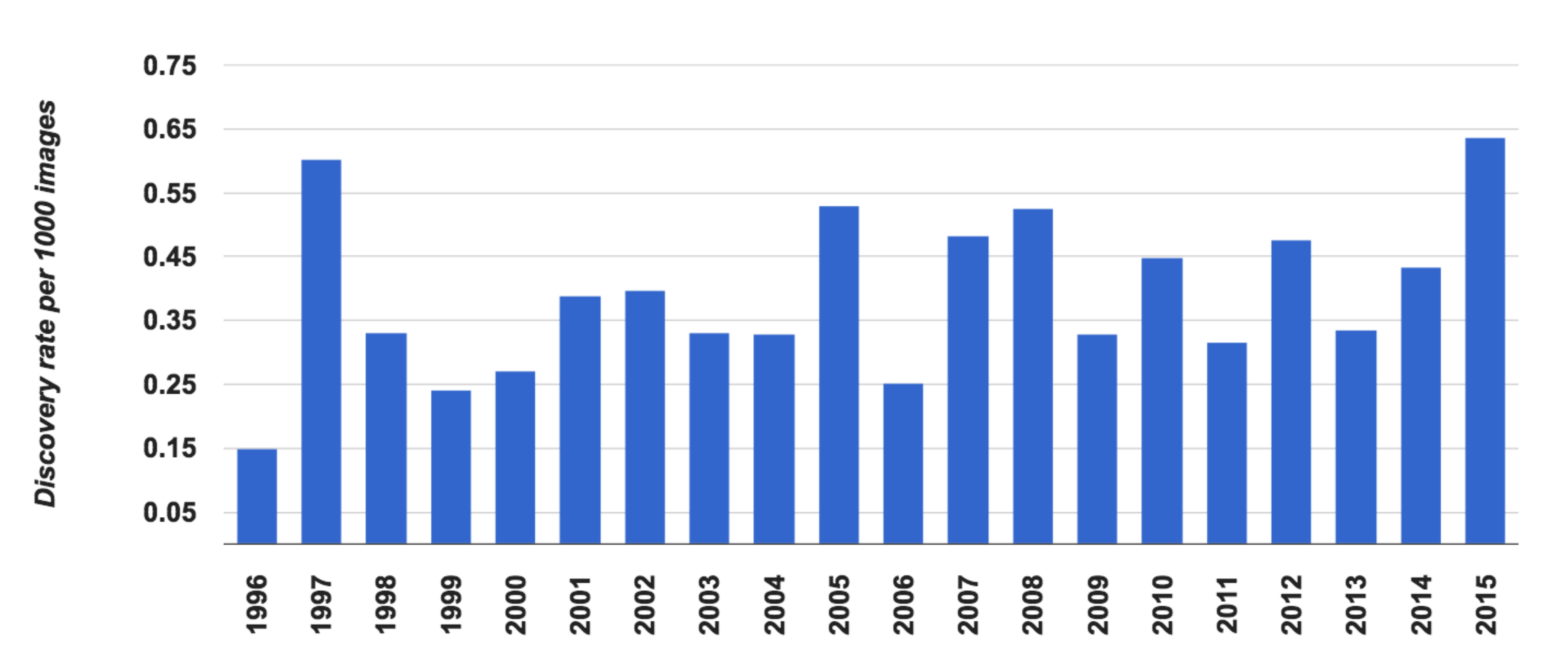}
  \caption[Images]{Number of Meyer-group discoveries per 1000 LASCO C2 images in each year during the period 1996--2015.}
\label{fig:meyer_rates}
\end{figure}

The Meyer-group discovery rate has remained largely commensurate with LASCO's data rate throughout the mission. Figure~\ref{fig:meyer_rates} shows the number of Meyer discoveries per year for every one thousand LASCO C2 images recorded. We do not include LASCO C3 observations here as only a limited number of very bright Meyer comets have been seen in C3, and all of those were also observed in C2. The somewhat anomalously low rate for 1996 is attributable to that being the first year of operation for the spacecraft, and its limited data return during the relatively Meyer-abundant months of January through March of that year. 

Based upon {\it SOHO}'s observations and the lack of any observations at larger heliocentric distances, the implication is that the Meyer group is an extremely old and highly-evolved population of largely inert bodies. While we can only speculate upon the origin, it would seem probable that the original fragmentation event leading to this group's formation occurred many revolutions ago. Assuming an orbital period of at least a century and many dozens of orbits needed to smooth out arrival times, this implies an origin at least 10$^4$ yr ago, perhaps substantially longer. Assuming the current rate of $\sim$10 comets/orbit holds all the way around the orbit and that Meyer comets are comparable in size to 322P/SOHO~1 (150--320 m in diameter; \citealt{knight16}), the original parent need only have been a few kilometers in diameter, e.g., comparable to typical Jupiter-family comets \citep{lamy04}. 

\subsection{96P Family Comets}
\label{sec:96p}
The Marsden and Kracht groups were originally recognized based on similarities in path across the {\it SOHO} fields of view of individual objects \citep{iauc7832, kracht02a}. Later, linkages between individual objects within each group were noted and each group's orbits were revised (e.g. \citealt{kracht02b}, \citealt{zhou05}, \citealt{zhou08}, \citealt{su08}, and many more), resulting in short period, low inclination orbits (see Table~\ref{table_overview}). Subsequent analysis by \citet{ohtsuka03} and  \citet{sekaninachodas04} showed that both the Marsden and Kracht groups are likely dynamically related to comet 96P/Machholz~1 as part of the larger ``Machholz Complex.'' According to these investigations, the Marsden and Kracht groups likely split from 96P some time in the last 800--1200 years and have been perturbed into their current orbits by having had slightly different gravitational interactions with Jupiter. Interestingly, the Marsden and Kracht groups currently have smaller perihelion distances and lower inclinations than 96P, but 96P was predicted to reach similarly extreme orbits in the next few centuries (e.g., \citealt{green90}, \citealt{mcintosh90}), well before the Marsden and Kracht groups were discovered.

Like the Meyer group, the Marsden and Kracht comets exhibit nearly identical condensed, stellar appearances (Figure~\ref{fig:non_krz_imgs}b\&c). All but one have been seen in C2, while the brightest are also seen in C3. Typical 96P family comets reach m7--8 but the brightest, Marsden comet P/1999~J6 = 2004~V9 = 2010~H3\footnote{It was also apparently recovered in September 2015 but this linkage has not yet been confirmed by the MPC.}, has reached $\sim$m5 each apparition \citep{lamy13}. The Marsden and Kracht comets are often observed to survive perihelion, and they generally peak in brightness within a few hours of perihelion. Their sizes are unknown, but they are likely larger than the Kreutz comets (e.g., at least a few 10s of meters in diameter) since there has been no appreciable fading of the objects definitively seen on multiple apparitions.

We do not show here the detection rates of 96P-family objects, as we feel it not particularly meaningful. These objects are highly dynamic, with certain individual members seemingly fragmenting between perihelion passages (e.g., \citealt{oates05}), and some objects not reappearing on subsequent passages. Furthermore, per the ``{\it SOHO} numbering" policy of the Sungrazer project, some objects are counted two or three times prior to determination of their linkage/periodicity, and then not counted on subsequent passages once they are deemed to be a ``known" object. While each group has seasonal variations in viewing geometry that affect detectability by {\it SOHO}, all Marsden and Kracht comets reach perihelion within the fields of view of the coronagraphs throughout the year. The relatively low number of family members likely dominates the seasonal and annual detection rates. 

Temporal clustering is frequent among the Marsden and Kracht comets, with two or more objects in the same family often arriving within days of each other followed by many months without any objects. This clustering has been interpreted as a sign of ongoing fragmentation (\citealt{sekanina05}; \citealt{knight08}\footnote{http://www.astro.umd.edu/people/Theses/2008knight.pdf}), with comets arriving in a cluster presumed to have split near perihelion on their previous perihelion passage. The most notable clustering event was the seven Kracht-group comets observed May 12-15, 2004 \citep{battams05}. \citet{knight08} concluded that the spread in orbital elements of each group is consistent with their having been produced by cascading fragmentation over the last few centuries. A ``family tree'' of each group (see, e.g., Figures~4.9 and 4.10 of \citealt{knight08}) can plausibly link the known members of each group to just a handful of fragments in the early 1990s. 

Despite the ongoing production of new group members via fragmentation, the overall number of comets in each of the two groups has remained relatively constant over $\sim$6 year intervals (the approximate orbital time for each group). This suggests that the smallest objects are dropping below the detection threshold and/or being destroyed. Thus, we are likely observing these groups at a fortuitous time; at some point in the relatively near future there are likely to not be any members of either group. It is possible that there are other groups in the Machholz complex that do not yet reach small enough perihelion distances to appear in {\it SOHO} images, but may evolve into the field of view in the future \citep{sekanina05}. 

The 96P family is evidently still evolving. During 96P's 2012 perihelion passage, two small fragments were observed on a similar trajectory and leading it by several hours. The fragments bore a remarkably similar appearance to members of the Kracht and Marsden groups, with no apparent coma or visible tail. Only one of the two fragments could be definitively linked to 96P \citep{battams13} due to the low quality astrometric reductions, however there is little doubt that both objects were indeed fragments of 96P. Both fragments appeared to survive perihelion and thus may be detectable during the 2017 perihelion passage, which will again be observed by {\it SOHO} assuming the satellite remains in operation.

\subsection{Non-Group}
We collectively refer to all of the comets that are not members of one of the previously discussed families as ``non-group.'' While this includes a handful of comets that are apparently dynamically related to each other, the majority of the ``non-group'' comets have no relation to any other known objects. These comets are a mixture of short and long period comets, and some have substantially larger perihelion distances than typical near-Sun comets. Due to the lack of relationships between the non-group comets, there are no group properties or discovery statistics to discuss. We briefly highlight below a few of the most interesting objects, and direct the reader to \citet{lamy13} for photometry of all non-group comets through 2008.

The first short period comet to be definitively identified in {\it SOHO} images was 322P/SOHO~1 (P/1999~R1), which was recognized by R. Kracht \citep{kracht02c} and the orbit definitively determined by \citet{hoenig06}. It has now been observed on five apparitions and tentatively linked to C/2002~R5, C/2008~L6, and C/2008~L7 as members of the ``Kracht~II'' group \citep{hammer02}. 322P recently became the only short period {\it SOHO}-discovered comet observed beyond the fields of view of solar observatories \citep{knight16}; its inactivity at these distances and peculiar properties have called into question whether it is of traditional cometary or asteroidal origin. 

Additional confirmed periodic non-group comets in the {\it SOHO} dataset include 321P/SOHO~3 (P/1997~J6) which has been seen on a {\it SOHO}-record six apparitions, 323P/SOHO~2 (1999~X3) which has been seen on four apparitions, and P/2003~T12 (SOHO~7) which has been seen three times over four apparitions (it went unobserved during its 2007 apparition due to poor observing geometry from {\it SOHO}). Despite its much larger perihelion distance than typical {\it SOHO}-discovered comets ($q$=0.57 AU), P/2013~T12 reached extraordinarily high phase angles in STEREO-B images during its 2012 apparition, which \citet{hui13} used to constrain dust scattering and polarization models. There have been very few orbits computed for non-group comets since 2010, making it possible that at least a few additional non-group comets have been re-observed but not yet recognized in the {\it SOHO} data. 

The majority of the non-group comets are on high inclination and apparently long period orbits. However, orbit determinations for these comets are highly ambiguous, often with numerous divergent, unique solutions. Most appear ``stellar'' and are only observed for a few hours, thus little is known about them besides their brightness (see \citealt{lamy13}). One prominent recent exception was C/2015 D1 (SOHO) which was observed for nearly four days in {\it SOHO} images. Although it was destroyed during the perihelion passage, enough dust survived that its remnants became the first {\it SOHO}-discovered sunskirting comet ever observed from the ground \citep{hui15}. 

While it was not discovered by {\it SOHO}, non-group comet C/2012~S1 (ISON) was the most well-studied sungrazing comet in history and, therefore, deserves a brief mention here. ISON was discovered more than a year before perihelion and gained widespread fame because it was the first known dynamically new (apparently entering the inner solar system for the first time) comet on a sungrazing orbit. The long lead time allowed it to be well characterized prior to reaching the {\it SOHO} fields of view. Unfortunately, ISON disintegrated prior to perihelion \citep{knight14,sekanina14} and failed to became the naked-eye comet many had hoped for. Nonetheless, it was observed far more extensively than any other near-Sun comet, resulting in a wealth of unique information. These include constraints on its nucleus size prior to entering the {\it SOHO} fields of view \citep{li13,delamere13}, production rates of water and other volatiles (e.g. \citealt{combi14}, \citealt{knight15}), and identification of outbursts of activity that may have resulted in the catastrophic breakup of the nucleus \citep{opitom13,schmidt15}.

\section{Conclusion}
Since its launch in 1996, {\it SOHO} continues to discover sungrazing and near-Sun comets at a consistent and high rate, enabled by the NASA-funded Sungrazer project. As of May 31, 2016, 3,138 objects have been discovered by the project, primarily (86\%) belonging to the Kreutz-group of Sungrazing comets, with the remaining objects being distributed among the Meyer group (6\%), the 96P Family (Marsden and Kracht groups; 3\%), and so-called non-group objects (5\%). In this paper we have provided a review of {\it SOHO}'s detections and observations of each of these populations. The rates of discovery of objects in the Kreutz and Meyer groups, when corrected for the data rate of the spacecraft, have remained largely consistent over at least the past ten to twelve years and suggest that the current discoveries are nearly complete and approaching the limits of instrumental capabilities. We discussed the diverse morphology of Kreutz group objects, recommending this as a focus area of future studies on the population, and showed examples of the rather uniform morphologies exhibited by all members of other groups. We also discussed the arrival rates of members of the Meyer, Marsden, and Kracht groups, showing that the Meyer group is consistent with an evolved population while the Marsden and Kracht groups continue to fragment. This ongoing evolution is not restricted to the small comets in the 96P family -- 96P itself was accompanied by two fragments unseen from the ground during its most recent apparition in 2012, highlighting the valuable role {\it SOHO} continues to play for solar system studies after 20 years.

\begin{figure}
  \centering
  \includegraphics[width=88mm]{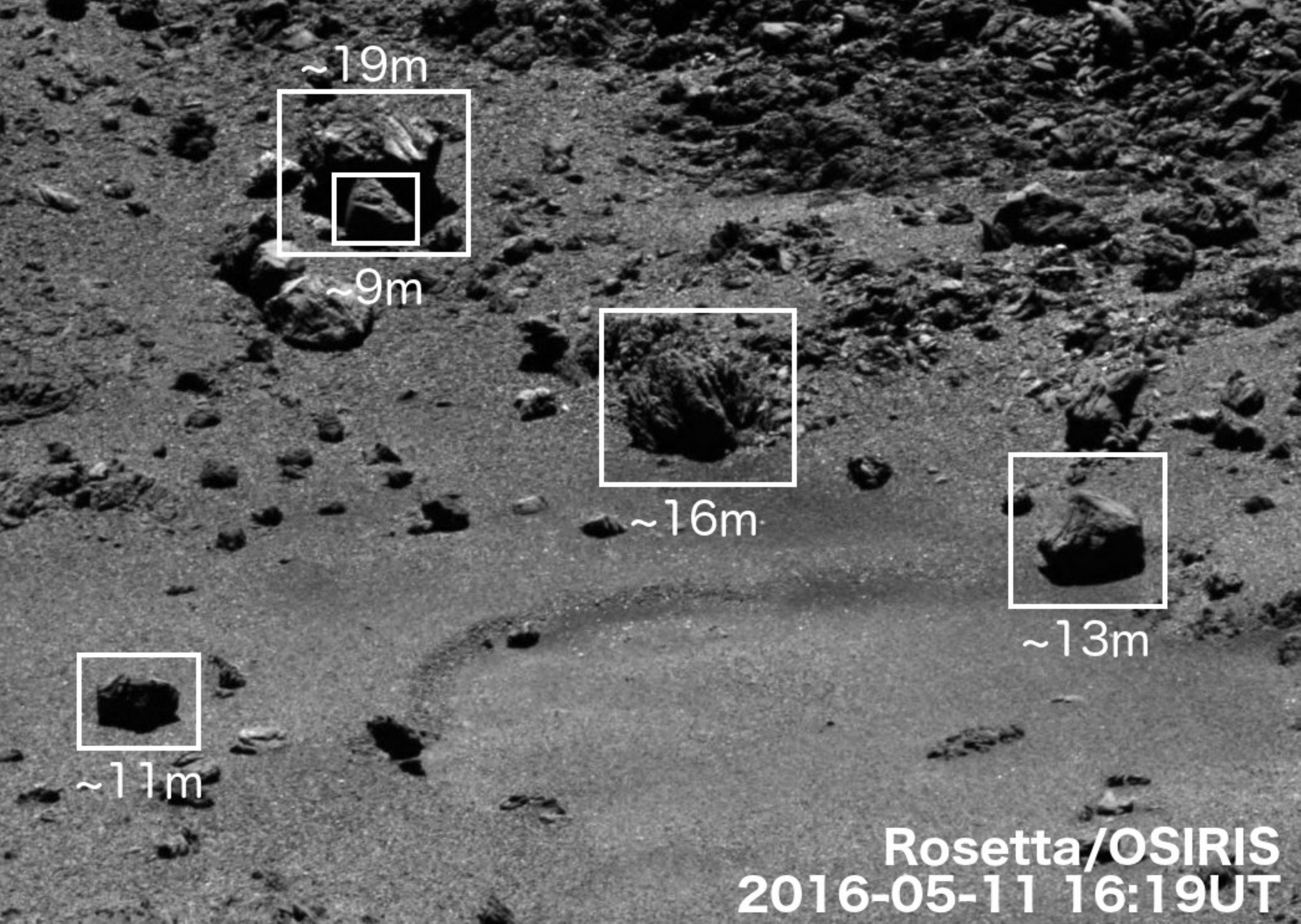}
  \caption[Images]{Annotated Rosetta/OSIRIS image highlighting large boulders on the surface of comet 67P with diameters similar to those estimated for {\it SOHO} comets. Image credit: ESA/Rosetta/MPS for OSIRIS Team MPS/UPD/LAM/IAA/SSO/INTA/UPM/DASP/IDA }
\label{fig:rosetta_boulders}
\end{figure}

We conclude with a thought experiment to highlight how the studies of small comets by {\it SOHO} complement work going on elsewhere in the Solar System and to link this manuscript with others in these proceedings. Rosetta observations of 67P/Churyumov-Gerasimenko have shown that, while a low level of activity seems to be nearly uniform across the daytime surface, strong outbursts are generally limited to vertical faces (\citealt{vincent16}; \citealt{farnham13} drew a similar conclusion from flyby data of 9P/Tempel 1) suggesting some localized differences in composition. Surface features are seen on 67P on many scales, from meters to hundreds of meters, that exhibit different physical properties. For example, Figure~\ref{fig:rosetta_boulders} shows a number of several-meter class boulders on the surface of 67P that, presumably, are somewhat less volatile than, say, source regions of jets. 

If we now place 67P in a hypothetical Kreutz-like orbit and fragment the comet, we could hypothesize the current 21 km$^3$ nucleus \cite{sierks15} to completely disrupt and form six 1.5-kilometer diameter objects, around 90 0.5-kilometer objects, and approximately one million twenty-meter class objects. The six large objects would be analogous to the historically brightest members of the group and the 90 intermediate sized objects would be analogous to C/2011 W3 Lovejoy. The smallest objects, about the size of boulders seen on the surface of 67P would be analogous to the $\sim$2,700 Kreutz comets discovered by {\it SOHO} thus far. Given the likely differences in volatile content between a relatively inert boulder on the surface of 67P and a volatile-rich boulder from its interior, it is likely that our hypothetical Kreutz population would exhibit different morphologies and brightening behaviors near the Sun. Thus, studies of the various aspects of sungrazing comets of all size can help to piece together the properties of the parent comet, allowing us glimpses of the interior of a comet not possible even with Rosetta. 

\smallskip
{\bf Data Accessibility. } SOHO/LASCO data are freely available from a searchable database hosted at http://lasco-www.nrl.navy.mil

\smallskip
{\bf Authors' Contributions.}  KB compiled the project statistics and compiled the plots and figures used in this paper. Both MMK and KB contributed equally to the analysis of the data and composition of the paper.

\smallskip
{\bf Competing Interests. } The author(s) declare that they have no competing interests.

\smallskip
{\bf Funding. } KB was supported by the NASA-funded Sungrazer Project. MMK was supported by NASA Outer Planets Research grant NNX13Al02G.
\smallskip

\section*{ACKNOWLEDGMENTS}
We thank the anonymous referee for a prompt review. This work benefited from discussions with the international team ``The Science of Near-Sun Comets'' led by G. Jones at ISSI (International Space Science Institute) in Bern, Switzerland, in 2014--2015. Both authors would like to acknowledge the enormous and invaluable efforts of the amateur astronomers who contribute their free time to search for and discover {\it SOHO}'s comets via the Sungrazer Project. Little of the work presented here, and in many of the references within, would be possible without their support and dedication to the Project.


\end{document}